# DSP-Enhanced OTDR for Detection and Estimation of Events in PONs

Manuel P. Fernández, Laureano A. Bulus Rossini, Juan Pablo Pascual and Pablo A. Costanzo Caso, *Member, IEEE.*

***Abstract*— To plan a rapid response and minimize operational costs, passive optical network operators require to automatically detect and identify faults that may occur in the optical distribution network. In this work, we present DSP-Enhanced OTDR, a novel methodology for remote fault analysis based on conventional optical time-domain reflectometry complemented with reference traces. Together with the mathematical formalism, we derive the detection tests that result to be uniformly most powerful, which are optimal according to the Neyman-Pearson criterion. To identify the type of fault and fully characterize it, the detection stage is followed by the estimation of its characteristic parameters, such as return loss and insertion loss. We experimentally demonstrate that this approach allows to detect faults inside the event dead zone, which overcomes the shortage of conventional event-marking algorithms. In our experiments, we achieved detection sensitivities higher than 0.2 dB in a 1:16 split-ratio PON, and higher than 1 dB in a 1:64 split-ratio PON, with estimation errors that can be as low as 0.01 dB. We also verified how the optical network terminal's reflectivity can improve the detection capabilities. This monitoring methodology involves negligible additional cost for the operator.**

## I. Introduction

During the last years, a wide amount of networks with fiber to the home (FTTH) architectures were deployed to satisfy the increasing customer's bandwidth demand. Among these networks, the different technologies of passive optical networks (PON) are extremely cost-efficient solutions to deliver a fiber connection to the users and they have been massively deployed [1]. Consequently, physical layer supervision of these type of networks has been widely studied. A complete monitoring system must accomplish several main objectives and requirements. Firstly, it must allow the automatic detection and localization of events that may occur in the optical distribution network (ODN) in favor of operational expenditures (OPEX) savings [2] by avoiding the need for in-field testing and diagnose. Moreover, it must timely detect and characterize different faults to plan a quick response, minimizing the mean down time (MDT), which has direct impact on the customers' satisfaction. In addition, it is extremely important that the monitoring solution involve minimum capital expenses (CAPEX) for the network operator and it must be easily installed in already deployed PONs [3]-[4].

Currently, the use of an optical time-domain reflectometer (OTDR) is recognized as being the most efficient technique to characterize an optical link [5]-[6]. However, the direct application of OTDR in point-to-multipoint networks such as PONs presents severe limitations: the addition of backscattered signals from different distribution drop fibers (DDF) makes the task of identifying the branch in which an event occurred very difficult to achieve. Since reflections arising from several branches are expected, the event dead zone becomes critical in this type of architectures. Also, the use of power splitters critically reduces the backscattered power and therefore the detection capabilities through conventional signal processing techniques are limited.

Several OTDR-based techniques have been proposed to overcome the aforementioned limitations without using active components in the ODN. The use of highly reflective filters in each DDF termination allows to localize events if the terminations are accurately separated [7]-[8]. Other approaches are based on the use of reference reflectors at different wavelengths and tunable OTDR [10]-[11]. Some authors demonstrated the use of tunable OTDR together with multiplexers as a passive bypass in the remote node in order to assign a monitoring wavelength to each branch [12] or group of branches (e.g. 8 branches) [13]. The use of encoders as demarcation devices has been demonstrated to complement the OTDR functionality [14]-[17]. All these techniques require to have additional components in the ODN, involving higher CAPEX and OPEX. Coherent detection-based OTDRs can achieve high dynamic ranges, but suffer severe penalties due to polarization issues, phase noise and coherent Rayleigh noise (CRN) [18]. Recently, the use of photon-counting OTDR has been demonstrated to The authors are with Laboratorio de Fotónica en Microondas y Comunicaciones Ópticas, Instituto Balseiro, Bariloche 8400 (RN), Argentina. M. P. Fernández is fellow of CONICET (email: manuel.fernandez@ib.edu.ar). P. A. Costanzo Caso (email: pcostanzo@ib.edu.ar), L. A. Bulus Rossini (email: lbulus@ib.edu.ar) and J. P. Pascual (email: juanpablo.pascual@ib.edu.ar) are researchers of CONICET and professors at Instituto Balseiro, UNCuyo-CNEA.

obtain both high sensitivity and resolution by using a single-photon detection scheme [19]-[20]. This is, however, a more complex and expensive OTDR structure that is not commercially available yet.

Although the mentioned proposals allow to remotely detect faults in several PON scenarios, it is extremely important and desirable from an operators' point of view to exploit the advantages of the remote monitoring with conventional OTDR and digital signal processing (DSP) techniques: simplicity, low-cost, transparency to data signals, scalability and easy implementation in already operative PONs. Hence, in this work we propose a novel methodology for OTDR-based fault analysis, which we called DSP-Enhanced OTDR (DSPE-OTDR). In this approach, an event detection algorithm based on DSP techniques is applied to the comparison of the acquired measurement with a reference one, obtained during the normal operation of the network. We derive the detection test that results to be the uniformly most powerful, which is optimal according to the Neyman-Pearson criterion. Expressions for the detection and false alarm probabilities together with the optimal decision thresholds are obtained. A comprehensive performance evaluation as a function of the OTDR characteristic parameters and the network topology is also carried out. The detection stage is followed by a maximum-likelihood estimation process to obtain the event parameters, which allows to remotely characterize and identify the type of fault.

Although previous works as [9], [13], [21] and [22] qualitatively describe monitoring techniques based on the use of reference measurements, this is the first time that these detection and estimation concepts, which had major success in modern radar systems and digital communication receivers, are proposed for monitoring optical networks.

Using this approach, we are able to detect, characterize and identify different types of faults, such as a link breaks, connector misalignments and fiber bendings, with high in PON architectures with split-ratios up to 1:64 within short measurement times. Even faults that lie within the dead zone are successfully detected, demonstrating that the algorithm is especially suitable for high-density PONs, where a great amount of dead zones are expected. We also verify how using the intrinsic reflectivity of the optical network terminal (ONT), the detection capabilities can be largely improved.

The present work is organized as follows. In Section II the fundamentals of DSPE-OTDR and the mathematical formalism for the acquired signals are introduced. From these models, in Section III the detection tests are derived and their performance is evaluated. Then, the estimator for the event parameters are found. Finally, the proof-of-concept experiments to demonstrate the algorithm's effectiveness are carried out in Section IV, emphasizing in situations where conventional event-marking algorithms fail, such as with small faults inside the event dead zone.

## II. Fundamentals of DSPE-OTDR for PON Monitoring

### A. *Description*

We assume a star topology for the PON, such as that illustrated in Fig. 1. The ODN is composed by a feeder fiber,

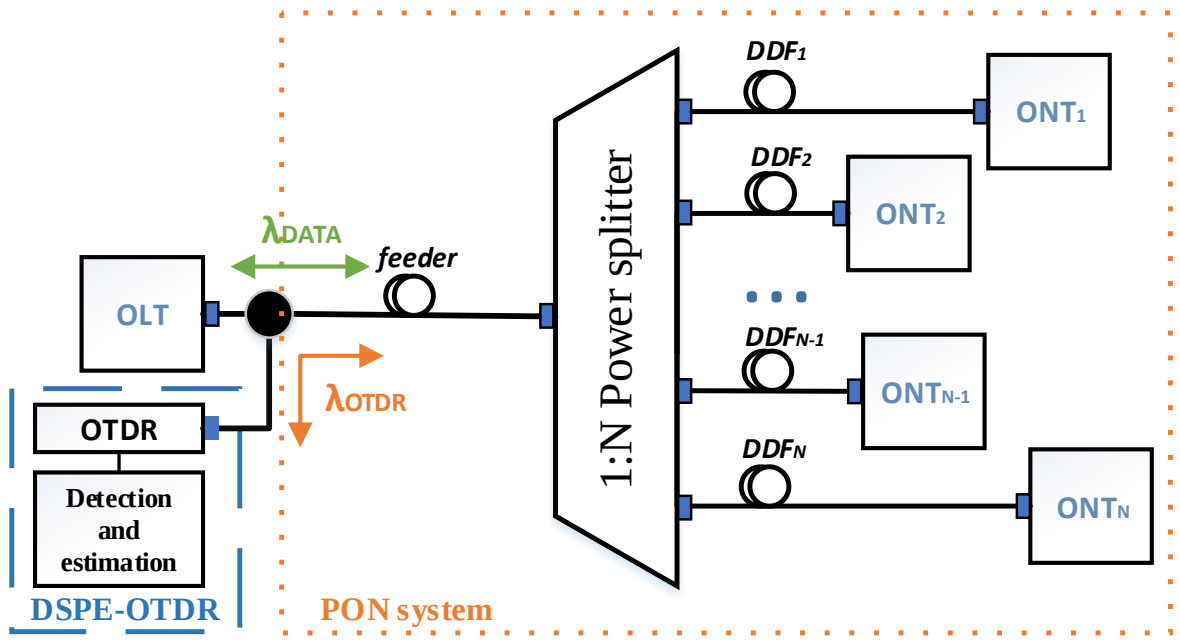

Fig. 1. Scheme of the OTDR-based remote monitoring system.

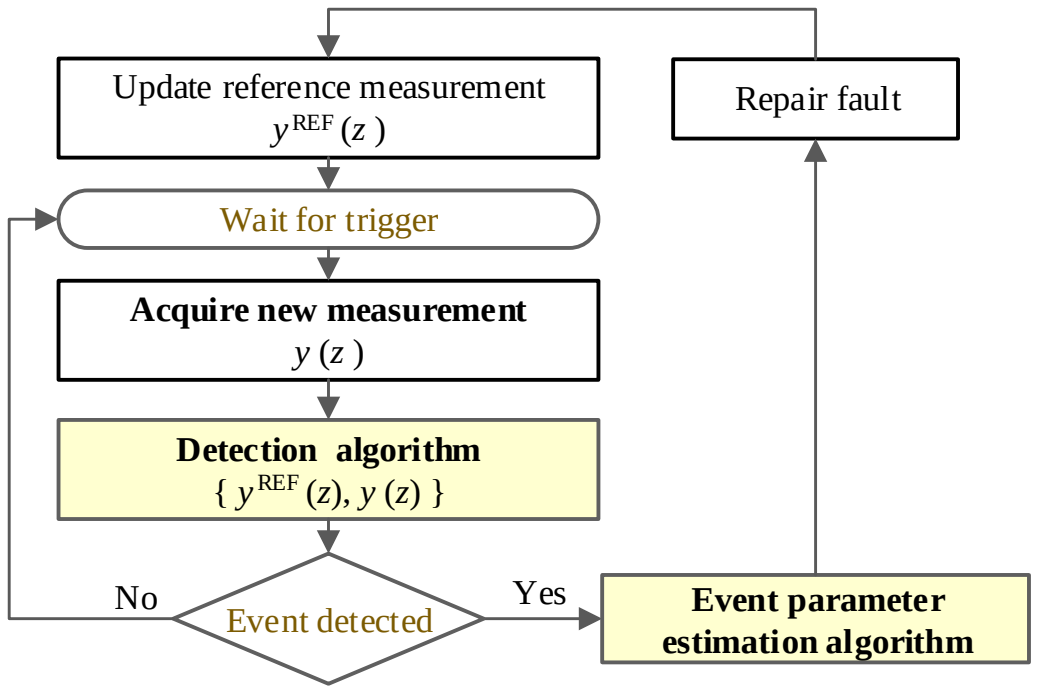

Fig. 2. Flow diagram of the DSPE-OTDR for PON monitoring.

a 1:$N$ power splitter and $N$ drop fibers connected to it, which derive the data signals to $N$ optical terminals. A conventional OTDR is connected to the PON through a wavelength multiplexer to be completely transparent to the data signals.

The flow diagram of the DSPE-OTDR-based remote monitoring scheme is depicted in Fig. 2. Previous to the operation, a reference measurement $y^{REF}(z)$, corresponding to the PON under normal condition, is obtained and stored in a local database. Ideally, this signal is either noise-free or presents a high SNR and can be obtained by averaging over a large number of measurements. This is reasonable and tolerable since the mean time between failures in this type of networks is relatively high.

During operation, the monitoring system stays in an idle state until a new measurement, which can be triggered periodically or manually, is performed. When this occurs, the OTDR acquires a new measurement using the dedicated monitoring wavelength. Once the current measurement $y(z)$ is obtained, the detection algorithm compares it with the reference signal, where the events will appear as deviations with respect to the reference signal.

In the proposed detection algorithm, events are detected through the occurrence of reflections and losses with respect to the reference signal. Specifically, a statistic, function of the acquired noisy measurement, is compared to a predefined threshold and two binary decisions are made:

(a) *there is* / *there is not* a reflection.
(b) *there is* / *there is not* a loss.

If an event is detected, the operator can use additional information, such as the location of each user's ONT reflection in the trace, to identify the faulty branch.

Subsequently, for the event to be completely characterized, its characteristic parameters, such as return loss and insertion loss, are estimated from the observation samples.

After the fault is repaired and previous to start again with the operation stage, the reference signal needs to be updated as explained before. It should be pointed out that the proposed algorithm is compatible with the use of conventional event-marking algorithms and it is completely scalable, since to add new users to the PON only requires updating the reference measurement.

### B. *Mathematical model for the reference signal.*

The OTDR signal will be expressed as a function of the distance $z$ (in km). Thus, the fiber parameters are the following: $K'$ is the fiber backscattering factor, expressed in terms of the spatial pulse width and is related to the backscattering factor $K$ through $K' \equiv 2K/v_g$, with $v_g$ being the group velocity in the fiber. The parameter $\alpha$ (in dB/km) is the attenuation constant of the fiber.

To obtain an algebraic expression for the detected OTDR signal at a given sample $z = z_i$ after the power splitter, we make the following assumptions:

- The probe pulse is a rectangular pulse of power $P_0$ and temporal width $T$, which corresponds to a spatial width of $W \equiv T v_g/2$.
- There are $N_a \leq N$ users whose ONT localization is larger than the distance corresponding to the observation sample $z_i$.
- The total round-trip insertion loss and power penalties at the distance $z_i$ in the reference measurement is the same for all the DDFs and are included in a parameter called $f_l$.

We focus on two cases depending on the observation sample. If there are no reflections, the reference signal is composed only by the superposition of backscattered power from $N_a$ drop fibers, so $y^{REF}(z)$ evaluated at the observation sample $z = z_i$ can be written as

$$y^{REF}(z_i) = f_l \frac{1}{N^2} N_a P_0 W K' 10^{-\frac{2\alpha z_i}{10}}. \quad (1)$$

The second case considered is when the observation sample corresponds to the reflection from the ONT reflective termination of the $e$-th branch, DDF$_e$, localized at $z = z_{\text{ONTe}}$. Assuming that the reflected power is much larger than the backscattered power of this DDF, then the reference signal at the observation sample $z_i \in (z_{\text{ONTe}} - W/2, z_{\text{ONTe}} + W/2)$ can be expressed as

$$y^{REF}(z_i) = f_l \frac{1}{N^2} P_0 \left( N_a W K' 10^{-\frac{2\alpha z_i}{10}} + 10^{-\frac{RL_{\text{ONTe}}}{10}} 10^{-\frac{2\alpha z_{\text{ONTe}}}{10}} \right), \quad (2)$$

where the first term accounts for backscattering contributions from $N_a$ branches and the second term is due to the Fresnel reflection of the ONT of the user $e$, which is characterized by a return loss $RL_{\text{ONTe}}$ (in dB).

### C. *Mathematical model for the signal after a fault.*

We next assume that a fault occurred in the DDF$_e$ at a distance $z_{\text{event}}$. This event is characterized by a return loss $RL_e$ and an insertion loss $IL_e$ (both expressed in dB). The noisy acquired signal $y(z)$ will show deviations with respect to the reference signal. Two classes of deviations are considered, namely reflections and losses.

#### *Reflections*

Let us consider that a reflective event, characterized by a return loss $RL_e$, produces a reflection in the detected signal at the sample $z_i \in (z_{\text{event}} - W/2, z_{\text{event}} + W/2)$. Assuming that the Fresnel-reflected power of the event is much larger than the backscattering level and that there was not a previous reflection at $z_i$, then the detected signal when a reflection occurs can be expressed in terms of the reference signal in Eq. (1) as

$$y^R(z_i) = y^{REF}(z_i) + f_l \frac{1}{N^2} P_0\, 10^{-\frac{2\alpha z_{\text{event}}}{10}}\, 10^{-\frac{RL_e}{10}}. \quad (3)$$

#### *Losses*

If the event is characterized by an insertion loss $IL_e$, the detected signal evaluated at a $z_i$ behind the loss inducing event, can be expressed in terms of the reference signal in Eq. (1) as

$$y^L(z_i) = y^{REF}(z_i) + f_l \frac{1}{N^2} P_0\, W\, K' 10^{-\frac{2\alpha z_i}{10}} \left( 10^{-\frac{2IL_e}{10}} - 1 \right), \quad (4)$$

where the term $10^{-\frac{2IL_e}{10}} - 1 < 0$.

As mentioned, the ONT reflective termination, localized at the interval $z_i \in (z_{\text{ONTe}} - W/2, z_{\text{ONTe}} + W/2)$, can be used to enhance the detection capabilities. As assumed for Eq. (2), the reflected power is much larger than the backscattered power. Then, the detected signal at the observation sample corresponding to the reflective termination of the DDF$_e$ can be written in terms of the reference signal in Eq. (2) as

$$y^L(z_i) = y^{REF}(z_i) + f_l \frac{1}{N^2} P_0 10^{-\frac{RL_{\text{ONTe}}}{10}} 10^{-\frac{2\alpha z_{\text{ONTe}}}{10}} \left( 10^{-\frac{2IL_e}{10}} - 1 \right). \quad (5)$$

### D. *Noise*

Each sample in the acquired signal contains a random noise term $y_N(z_i)$ whose statistics define the detection approach. Commercial OTDRs use laser diodes that have linewidths of several nanometers, which leads to low coherence lengths compared to the pulse width, and consequently, CRN is negligible. For example, a laser source with a spectral width of 20 nm centered at 1550 nm has a coherence length of $c/(\pi\Delta\nu) = 38.2$ µm, which is much smaller than the common pulse width of several meters.

The detected power is normally very low, thus the detector operates in the thermal noise limit. Additionally, since several hundreds of averages are performed, the central limit theorem states that it is reasonable to assume that the noise follows a Gaussian distribution. Therefore, the resulting noise term $y_N(z_i)$ can be accurately modeled as an additive Gaussian random variable with zero-mean and variance $\sigma_N^2$ [5], [24].

It should be remarked that the detection algorithm described here is aimed to conventional OTDR devices. For other OTDR configurations, their intrinsic noise statistics, such as CRN in coherent-OTDR and the Poisson noise in photon-counting OTDR [20] should be taken into account.

## III. DSPE-OTDR: EVENT DETECTION AND PARAMETER ESTIMATION

### A. Optimal Detection Algorithm

The OTDR transmits the probe pulses and then samples and averages the photodetected signals, where each sample of the averaged signal $y(z_i)$ is a random variable. The detection procedure is then divided into two hypothesis tests:

(a) *Decide between $H_R$ and $H_0$.*
(b) *Decide between $H_L$ and $H_0$.*

Under the null hypothesis, $H_0$, it is assumed that neither reflection nor loss are present with respect to the reference measurement. Under the reflection hypothesis, $H_R$, it is assumed that the measurement presents a reflection. Finally, under the loss hypothesis, $H_L$, the observation sample experiences a loss with respect to the reference. Since the noise is Gaussian with zero-mean and variance $\sigma_N^2$, for all the hypotheses, $y(z_i)$ follow a Gaussian distribution with the same variance and different mean values. The hypotheses can be summarized as

$$\begin{aligned} \boldsymbol{H_0} &\rightarrow y(z_i) \sim \mathcal{N}(\mu_0(z_i), \sigma_N^2) \\ \boldsymbol{H_k} &\rightarrow y(z_i) \sim \mathcal{N}(\mu_k(z_i), \sigma_N^2) \end{aligned} \tag{6}$$

where $k = R, L$ stands for the reflection hypothesis and the loss hypothesis, respectively. The mean value of the hypotheses for each sample are obtained from Eqs. (1)-(5) as

$$\begin{aligned} \mu_0(z_i) &= y^{REF}(z_i) \\ \mu_R(z_i) &= y^R(z_i) \\ \mu_L(z_i) &= y^L(z_i)\,. \end{aligned} \tag{7}$$

To design the detection tests we follow the well-known Neyman–Pearson criterion, that is, the decision is designed to maximize the probability of detection, $P_D^k$, under the constraint that the probability of false alarm, $P_{FA}^k$, does not exceed a given value. Although the problem has unknown parameters, such as $RL_e$ and $IL_e$, it is derived assuming these are known in order to get the uniformly most powerful (UMP) test [28]. The solution to this problem leads to the comparison of the likelihood ratio with a threshold $\gamma$ [29]

$$\frac{f_Y(y|H_0)}{f_Y(y|H_k)} \underset{H_k}{\overset{H_0}{\gtrless}} \gamma \tag{8}$$

where $f_Y(y|H_0)$ and $f_Y(y|H_k)$ are the probability density functions (PDF) of $y(z_i)$ given that the null hypothesis is true and given that the $k$ hypothesis is true, respectively. These functions are given by Eq. (6).

By replacing the expression for the PDFs in Eq. (8), after some algebraic manipulations and taking into account that $\mu_0 < \mu_R$ and $\mu_0 > \mu_L$, we obtain the well-known decision rules

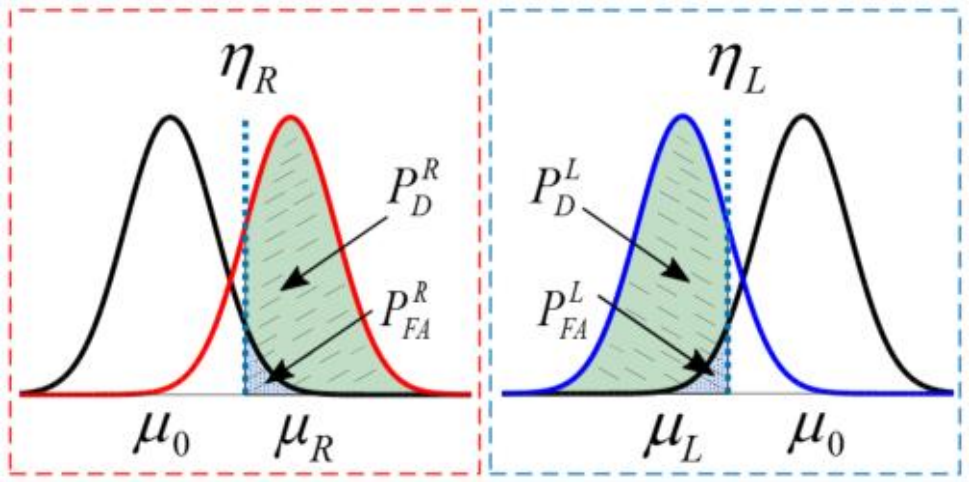

Fig. 3. Decision regions and probabilities for the two binary hypotheses tests.

$$y(z_i) \underset{H_0}{\overset{H_R}{\gtrless}} \eta_R(z_i) \qquad y(z_i) \underset{H_L}{\overset{H_0}{\gtrless}} \eta_L(z_i)\,, \tag{9}$$

where the sample $y(z_i)$ is the sufficient statistic that must be compared with $\eta_R$ for the reflection test and with $\eta_L$ for the loss test. Note the extremely low computational cost required, since the acquired signal must only be compared with a couple of predefined thresholds. Thus, it can be easily implemented in a dedicated unit or in the OTDR device.

The way to evaluate the performance of a detector is by means of its false alarm and detection probabilities. In our case these parameters are important since a wrongly claim that a fault occurred will cause undesired OPEX, while missing to detect a fault (e.g. a fiber break) will delay the restoration tasks, influencing on the customers' satisfaction.

Once an expression for $P_{FA}^k$ in terms of $\eta_k$ is derived, it is inverted to obtain the threshold by setting it in terms of a fixed value of $P_{FA}^k$. Since $y(z_i)$ is Gaussian distributed in both tests, the $P_{FA}^k$ of the reflection and loss tests are respectively

$$P_{FA}^R = \int_{\eta_R}^{\infty} \frac{1}{\sqrt{2\pi}\sigma_N} \exp\left(-\frac{(y-\mu_0)^2}{2\sigma_N^2}\right) dy = Q\left(\frac{\eta_R-\mu_0}{\sigma_N}\right) \tag{10}$$

$$P_{FA}^L = \int_{-\infty}^{\eta_L} \frac{1}{\sqrt{2\pi}\sigma_N} \exp\left(-\frac{(y-\mu_0)^2}{2\sigma_N^2}\right) dy = 1 - Q\left(\frac{\eta_L-\mu_0}{\sigma_N}\right), \tag{11}$$

where $Q(x)$ is the q-function [28].

It can be seen from Eq. (10)-(11) that the tests do not involve any unknown parameter and that the optimal decision thresholds only depend on the known parameters $\mu_0$, i.e. the reference signal, and on $\sigma_N$.

On the other hand, the probabilities of detection $P_D^k$ for the reflection and loss tests are respectively

$$P_D^R = \int_{\eta_R}^{\infty} \frac{1}{\sqrt{2\pi}\sigma_N} \exp\left(-\frac{(y-\mu_R)^2}{2\sigma_N^2}\right) dy = Q\left(\frac{\eta_R-\mu_R}{\sigma_N}\right) \tag{12}$$

$$P_D^L = \int_{-\infty}^{\eta_L} \frac{1}{\sqrt{2\pi}\sigma_N} \exp\left(-\frac{(y-\mu_L)^2}{2\sigma_N^2}\right) dy = 1 - Q\left(\frac{\eta_L-\mu_L}{\sigma_N}\right). \tag{13}$$

Figure 3 shows the tests decision regions together with their detection and false alarm probabilities. Since the noise variance is the same for all hypotheses, the detection probabilities only depend on the distance between the means of the hypotheses in each test. In the case of the loss detection, if the observation sample corresponds to the reflective termination, $P_D^L$ will be enhanced with respect to the observation at a sample $z_i$ outside the reflective termination if the condition

$$W\,K' 10^{-\frac{2\alpha z_i}{10}} < 10^{-\frac{2\alpha z_{\text{ONTe}}}{10}} 10^{-\frac{RL_{\text{ONTe}}}{10}} \qquad (14)$$

is verified, that is, if the ONT reflected power is larger than the backscattered power at the observation sample.

## B. Detection performance

In order to generalize the performance analysis, it is convenient to combine all the loss mechanisms to the observation sample $z_i$ in the acquired signal by defining the optical path loss ($OPL$) to $z_i$ at the monitoring wavelength as

$$OPL^2(z_i) = f_l \frac{1}{N^2} 10^{-\frac{2\alpha z_i}{10}}, \qquad (15)$$

which is reasonably assumed to be close to the optical path loss at the data wavelength. Recall that the maximum allowed optical path loss in the next-generation PON standards is 35 dB [30].

The most relevant figure of merit of commercial OTDR devices is the dynamic range, defined as the ratio between the backscattered power at the front-panel connector and the root mean square of the noise. Consequently, it is useful to relate the OTDR and fiber characteristic parameters to the dynamic range ($DR$) through the relation

$$DR = \frac{P_0 K' W}{\sigma_N}, \qquad (16)$$

which is normally given in dB units, taking 5log($DR$). It is well known that the dynamic range results from a trade-off between spatial resolution and acquisition time. Since the detected power from Fresnel reflections do not depend on the pulse width, in the following analysis, $DR$ is assumed to be specified for a pulse of $T = 100$ ns ($W \approx 10$ m) and a backscattering factor of $K = -82$ dB.

The means of the hypotheses $\mu_0$, $\mu_R$ and $\mu_L$ can therefore be written in terms of $OPL$ in Eq. (15) and $DR$ in Eq. (16), and the detection capabilities of the algorithm can be analytically evaluated from the OTDR specifications. More specifically, it is of special interest to determine the maximum $OPL$ that an OTDR with a dynamic range $DR$ can achieve, for a fixed $P_D^k$ and $P_{FA}^k$. Since in this type of detection systems, the false alarm and detection probabilities are critical, the former should be kept as low as possible and the latter should ideally be close to 1. In the following analysis, we fix $P_{FA}^k = 10^{-4}$ for both hypotheses tests, and we establish as a desirable design criteria to have a probability of detection $P_D^k > 0.95$ for the given $P_{FA}^k$.

In Fig. 4(a) the achievable $OPL$ is represented as a function of the OTDR dynamic range in the case of the Reflection Detection for different values of $RL_e$. It can be observed that for $DR = 22$ dB, the detection of a reflection with $RL_e = 30$ dB can achieve the desired $P_D^R$ even in the most pessimistic scenario, that is, when $OPL = 35$ dB. For higher return losses such as $RL_e = 50$ dB and the same $DR$, the achievable $OPL$ is 24 dB. In the case of the Loss Detection in the backscattered signal, the maximum $OPL$ is represented in Fig. 4(b) (in solid line). To detect losses with a sensitivity of 0.1 dB in a PON with a maximum $OPL = 15$ dB (which allows a split-ratio up to 1:16), a dynamic range of at least 25.3 dB is required. In the same scenario, to detect a 3 dB loss event requires a dynamic range of 19.2 dB. One can immediately note that even strong losses demand relatively high dynamic ranges to be detected. This requirement can be relaxed when the ONT reflection is considered. In the same Figure, the improvement in the detection capabilities using the ONT reflection is depicted (in dashed lines), when assuming $RL_{\text{ONTe}} = 40$ dB. It can be observed that the maximum $OPL$ is increased about 11 dB. In this case, an OTDR having $DR = 30$ dB (for the given $T$) could achieve a 0.1 dB loss sensitivity in scenarios with an $OPL$ of up to 30 dB, such as PON with a 1:128 split-ratio and a reach of 20 km.

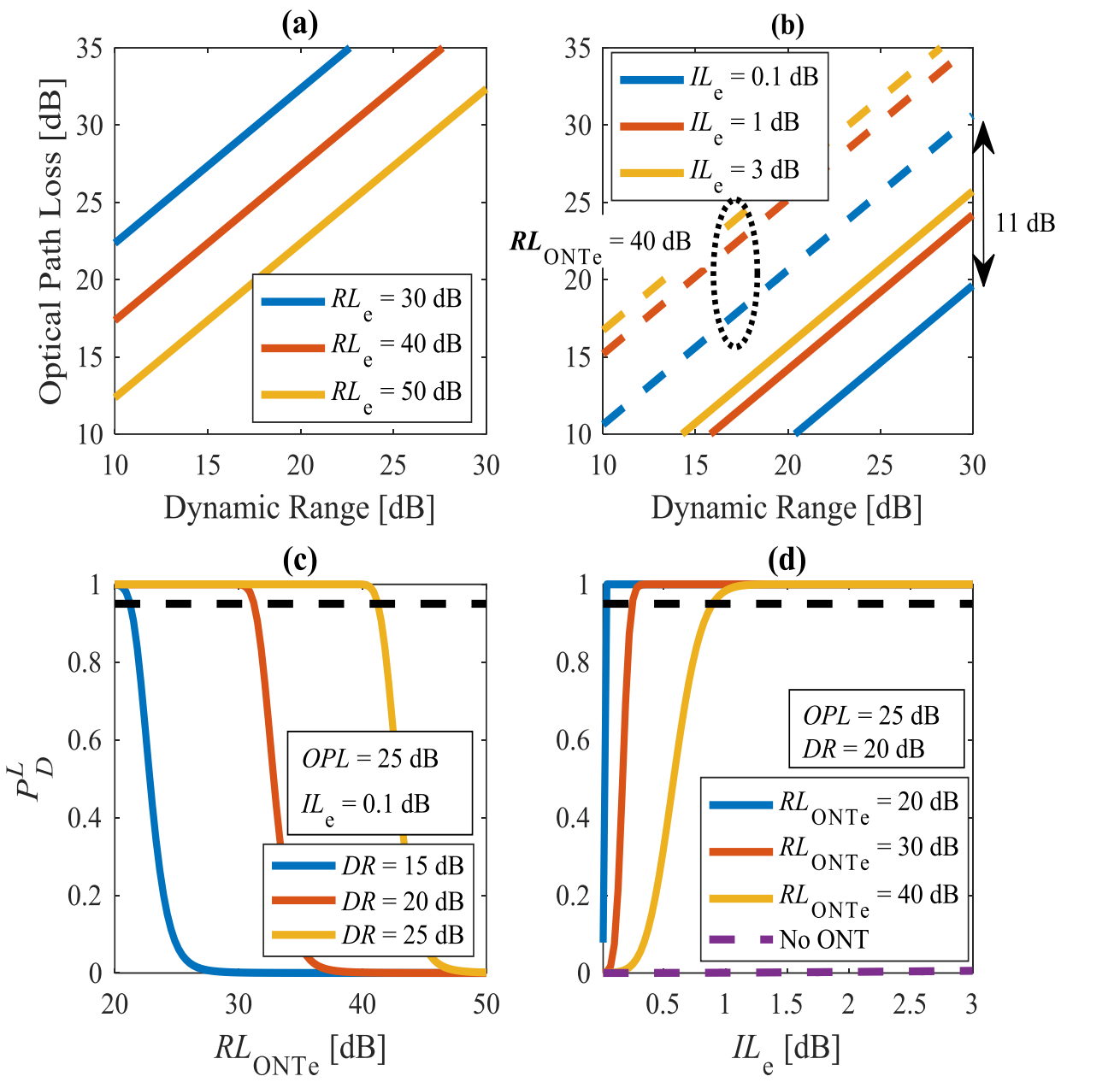

Fig. 4. Maximum achievable $OPL$ versus $DR$ to detect (a) reflections and (b) losses where it was fixed $P_{FA} = 10^{-4}$ and $P_D = 0.95$. Probability of loss detection versus (c) the ONT return loss and (d) the magnitude of the loss-inducing event.

It is clear from the previous analysis that the ONT reflective termination significantly improves the detection sensitivity. Thus, in Fig. 4(c), $P_D^L$ is represented as a function of $RL_{\text{ONTe}}$. Here, the event insertion loss is 0.1 dB and it is assumed that the path loss to the ONT is 25 dB, which is compatible with a 1:64 split-ratio and a reach of 20 km. To achieve the requirement of $P_D^L > 0.95$ with a $DR = 25$ dB, the ONT return loss should be lower than 41 dB, while highly reflective terminals with $RL_{\text{ONTe}} < 22$ dB accomplish this requirement even if $DR$ is as low as 15 dB. The sensitivity of the algorithm to the loss magnitude is analyzed in Fig. 4(d), where $P_D^L$ is depicted as a function of $IL_e$. For typical values of $RL_{ONTe} = 40$ dB, the sensitivity could be as high as 1 dB for an $OPL = 25$ dB, when using an OTDR with $DR = 20$ dB.

### C. Event parameter estimation

While in the detection stage of the algorithm, deviations with respect to the reference signal due to the occurrence of an event are detected, it is still necessary to find the characteristic parameters of the event for this to be completely characterized. Therefore, $RL_e$ and $IL_e$, which are deterministic values, must be estimated from the samples where reflections and losses were detected, respectively. To do this, we will follow the method of maximum likelihood.

Let us assume that either a reflection or a loss was detected at the samples $y(z_m)$, with $m = 1, \ldots, M$. If the noise is assumed independent and identically distributed, the $M$ observation samples are statistically independent random variables. Thus, the likelihood function under the hypothesis $k$ can be written as the product of the marginal PDFs, given by Eq. (6), and the estimation problem may be written as

$$\hat{\theta}_k = \arg\max_{\theta_k} \ell(\theta_k), \tag{17}$$

where $\hat{\theta}_k$ is the estimator of $RL_e$ and $IL_e$ when $k = R, L$, respectively, and

$$\ell(\theta_k) = -\frac{1}{2\sigma_N^2}\sum_{m=1}^{M}\left(y(z_m) - \mu_k(z_m)\right)^2. \tag{18}$$

is the log-likelihood function under the hypothesis $k$, where $\mu_k$ is function of the parameter $\theta_k$. In Eq. (18), constants that are not involved on the estimation problem were omitted. By solving the optimization problem of Eq. (17), the maximum likelihood estimator (MLE) for the return loss $\widehat{RL}_e$ and the insertion loss $\widehat{IL}_e$ can be found to be, respectively

$$\widehat{RL}_e = -10\log\left(\frac{\sum_{m=1}^{M}\left(y(z_m) - y^{REF}(z_m)\right)}{MP_0 OPL^2(z_{\text{event}})}\right), \tag{19}$$

$$\widehat{IL}_e = -5\log\left(\frac{\sum_{m=1}^{M} OPL^2(z_m)\left(y(z_m) - y^{REF}(z_m)\right)}{P_0 W K' \sum_{m=1}^{M} OPL^4(z_m)} + 1\right), \tag{20}$$

where the last applies only for the samples outside the reflective termination of the faulty branch. In the case of the observation samples corresponding to the reflective termination, the MLE for the insertion loss is given by

$$\widehat{IL}_e = -5\log\left(\frac{\sum_{m=1}^{M}\left(y(z_m) - y^{REF}(z_m)\right)}{MP_0 OPL^2(z_{\text{ONTe}})10^{-\frac{RL_{\text{ONTe}}}{10}}} + 1\right). \tag{21}$$

It can be verified that the estimators in Eq. (19)-(21) are consistent, i.e. they converge in probability to their true values as the number of observation samples $M$ increases.

In the case of the insertion loss, an accurate estimate of $IL_e$ is of extremely importance since in most common types of faults, this parameter is wavelength-dependent. Therefore, by testing at two or more different wavelengths, not only the nature (either reflective of non-reflective) and magnitude of the event can be obtained, but also the type of fault, e.g. a connector misalignment or a fiber bending, can be remotely identified.

TABLE I
MEASURED FIBER AND OTDR PARAMETERS

| Fiber parameter | Symbol | Value |
|---|---|---|
| Backscattering coefficient (for $T = 1$ ns) | $K$ | $-82$ [$dB$] |
| Attenuation coefficient | $\alpha$ | 0.21 [dB/km] |
| Group index | $n_g$ | 1.46 |
| **OTDR parameter** | **Symbol** | **Value** |
| Pulse peak power | $P_0$ | 31.8 [mW] |
| Source linewidth | $\Delta\lambda$ | 23 [nm] |
| Receiver bandwidth | $B$ | 9 [MHz] |
| Dynamic range (for $T = 100$ ns) | $DR$ | Averaging: 1 min. 19.16 [dB] Averaging: 3 min. 20.96 [dB] |

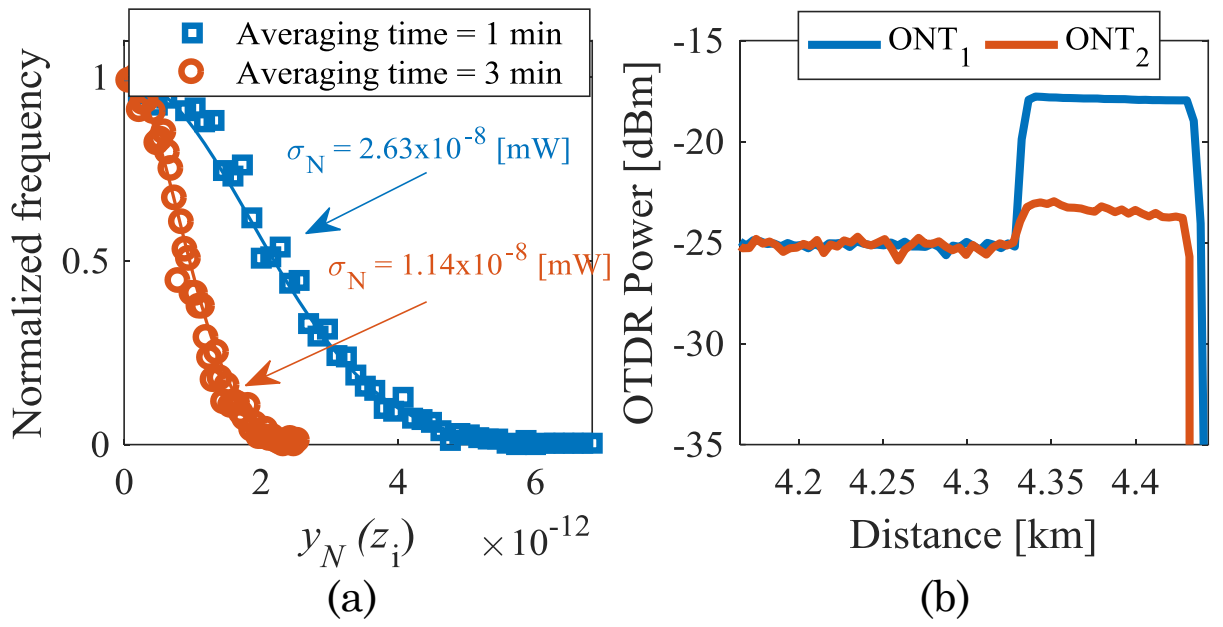

Fig. 5. (a) Histogram of OTDR noise for averaging times of 1 and 3 minutes, and (c) OTDR measurement of two ONT terminations.

## IV. Proof of Concept and Experiments

The proposed DSPE-OTDR was experimentally probed using a commercially available OTDR equipped with two laser sources at 1310 nm and 1550 nm. The test-bed PON deployed for the experiments uses single-mode fibers whose parameters at 1550 nm are shown in Table I.

### A. OTDR and ONT characterization

The measured characteristic parameters at 1550 nm from the OTDR used for the experiments are listed in Table I. The peak power of the probe pulses was found to be $P_0 = 31.8$ mW. The noise standard deviation $\sigma_N$ depends on the number of averages performed or, equivalently, the OTDR acquisition time. To estimate the standard deviation for a given OTDR configuration, we computed the histogram of acquired noise samples during acquisition times of 1 and 3 minutes when the OTDR measurement range was fixed at 25 km. Then, we fitted the one-sided histograms with Gaussian functions by means of a nonlinear least square method, as shown in Fig. 5(a). In both cases, a high coefficient of determination of about 0.99 is obtained. From this, the dynamic range of the OTDR corresponding to the different averaging times can be obtained by means of Eq. (16) and they are expressed in Table I for pulses of 100 ns. At 1550 nm the linewidth of the laser source is 23 nm, yielding a sub-millimeter coherence length and negligible CRN. Assuming propagation distances over a standard single-mode fiber of about 40 km, the widening due to dispersion for long probe pulses (>100 ns) is less than 2%. Finally, the bandwidth of the OTDR was found to be about 9 MHz.

All the previous characterization procedure can be equally extended to any commercially available OTDR device and to other monitoring wavelengths, as the U-band (1625-1675 nm), which is recommended for in-service monitoring tasks in optical access networks [25].

As mentioned, the intrinsic reflective characteristic of the users' ONT can be used to identify the termination point and to improve the detection capabilities if the condition in Eq. (14) is met. Moreover, the acquisition parameters must be chosen in a way that the backscattering floor is lower than the ONT reflected power. Thus, it is necessary to characterize the ONT in terms of its return loss at the monitoring wavelength. In Fig. 5(b) it is shown the acquired OTDR traces at 1550 nm, where it is seen the reflection obtained from two ONT modules. The first one, $ONT_1$, presents a lower return loss (higher reflectivity) of $RL_{ONT1} = 37.6$ dB, while the second, $ONT_2$, has a higher return loss (lower reflectivity) of $RL_{ONT2} = 49.4$ dB.

### B. *Event detection and parameter estimation*

To assess the performance of the proposed DPSE-OTDR, a test-bed PON was deployed. The network is composed by a feeder fiber of 2.7 km, a 1:$N$ power splitter and two branches are connected to it. The length of the drop fibers are $\{l_{DDFe} = 6.2\text{ km}, \ l_2 = 2.93\text{ km}\}$, where the $DDF_e$ is composed by two fiber spools of 2.95 km and 3.2 km, joined by a LC connector, and it is terminated with the previously characterized $ONT_1$. Previous to the operation, a reference trace for each test-bed PON was obtained. The network topology was deliberately chosen in a way that the induced events, which are small in magnitude, lie within a dead zone, thus emulating pessimistic detection conditions.

To choose the detection thresholds and the measurement parameters, such as pulse width and acquisition time, we set as a criterion that losses with a sensitivity of 1 dB must be detected with a $P_D \geq 0.95$, given that $P_{FA} \leq 10^{-4}$. It is important to point out that the sensitivity can be arbitrarily increased, as we will see later, by properly choosing the OTDR acquisition parameters. Figure 6 shows the result of applying the DSPE-OTDR to different fault scenarios. Together with the current measurement, it is shown the reference trace (in dashed lines) and the samples where reflections (red dots) and losses (black dots) were detected.

In the first experiment, the split-ratio of the PON is 1:32. We can then resort to the analysis in Section III.C to find the acquisition parameters that allow to accomplish the desired sensitivity. In the current scenario, the maximum $OPL$ is composed by the splitting loss (~15 dB), the maximum propagation loss (~2.1 dB) and the overall insertion loss (~2 dB), which leads to $OPL = 19.1$ dB. From Fig. 4(b) it can be obtained that a dynamic range of $DR = 24.9$ dB is required. In our OTDR, this $DR$ can be achieved, for example, using pulses of 500 ns and averaging over 3 minutes. Recall that shorter pulses, which lead to higher spatial resolutions, could be also used together with larger averaging times.

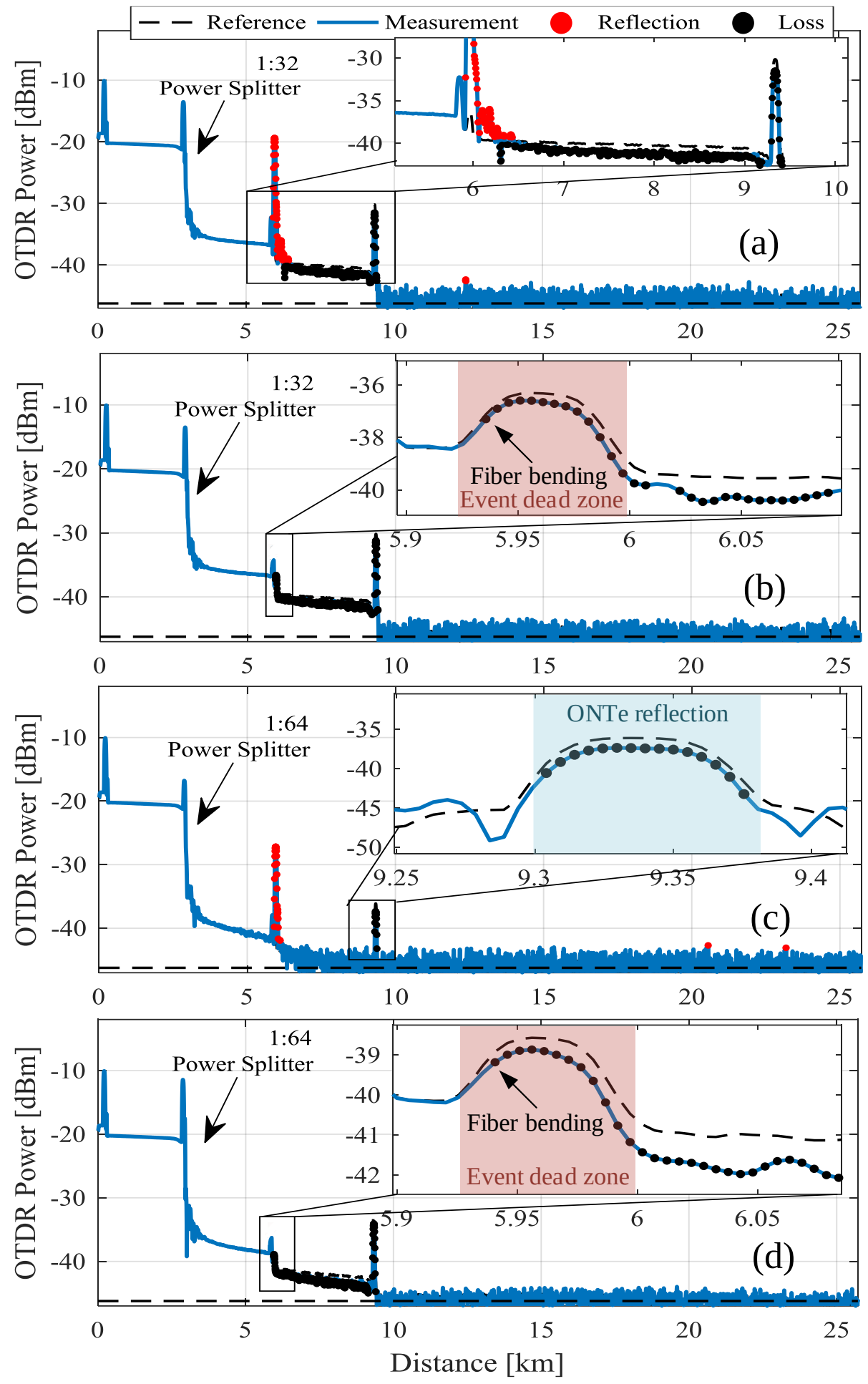

Fig. 6. Application of DSPE-OTDR in different faulty PON scenarios: (a) a connector misalignment in a 1:32 splitter, (b) a fiber bending in a 1:32 splitter, (c) a connector misalignment in a 1:64 splitter and (d) a fiber bending in a 1:64 splitter.

In the first place, a misalignment in the LC connector of the $DDF_e$ inducing an insertion loss of $IL_e = 1.2$ dB was generated. Figure 6(a) shows the detection result of the DSPE-OTDR, where it can be seen that a strong reflection at the connector's location (at 5.7 km) is detected. The power loss induced by the event is also clearly detected in the waveform from the event to the $ONT_e$ localization. The estimation algorithm was subsequently applied and the event parameters at 1550 nm were found to be $\widehat{RL}_e = 17$ dB and $\widehat{IL}_e = 1.1$ dB.

Under the same scenario, a small fiber bending was generated a few centimeters after the connector of the $DDF_e$. Hence, the fiber bending lies in a dead zone and consequently it is undetected by the OTDR's own event-marking algorithm. However, by applying the proposed detection algorithm, this non-reflective event is accurately detected, as it can be seen in the experimentally obtained waveform in Fig 6(b). In this case, the estimated value of $\widehat{IL}_e$ is found to be 0.94 dB.

The split ratio of the PON was next increased to 1:64 while the OTDR pulse width and acquisition times were kept fixed and the same faults were generated. In the case of the connector misalignment, the event reflection is still clearly identified, but the dynamic range is not high enough to accurately detect the losses in the backscattered signal. However, as shown in Fig. 6(c), the loss is still detected at the $ONT_e$ reflective termination and an accurate event insertion loss of 1.18 dB was obtained by means of Eq. (21). Note that in this case, the fault can be localized within the PON due to the reflective nature of the fault.

In the same scenario, in the case of a low-loss non-reflective fault, such as a fiber bending, the loss is still detected at the reflective termination, and the faulty branch can be identified. However, in order to localize such event within the PON, the OTDR measurement parameters should be adjusted in order to meet the requirements for the dynamic range. For instance, the maximum path loss in this scenario is $OPL = 22.2$ dB and hence, the required dynamic range to detect a loss of 1 dB in the backscattered signal, according to Fig. 4(b), is $DR = 27.9$ dB. In this example, to achieve this dynamic range, we kept fixed the pulse width and increased the averaging time to 8 minutes. The result of the application of the DSPE-OTDR to this fault scenario can be seen in Fig. 6(d), where the bending loss is accurately detected and it is estimated to have $\widehat{IL}_e = 0.97$ dB.

From the previous examples, it is clear how the proposed method can overcome the shortages of classical OTDR event-marking algorithms, providing a dead zone free automatic event detection and accurate event parameter estimation, even if small non-reflective faults are considered.

In the following, we show how using a dual-wavelength measurement, the type of fault can also be remotely identified from the estimated event parameters at each wavelength, since in common faults these are normally wavelength-sensitive. To exemplify this, three common types of faults are considered: a link break, a connector misalignment providing finite insertion loss and a fiber bending. It is well known that in the case of a fiber/connector break, a high reflection and infinite insertion loss are induced. The wavelength dependence of the mode-field diameter leads to a larger insertion loss at shorter wavelengths in a connector misalignment. On the other hand, the effective index in a fiber bending produces higher losses at longer wavelengths and negligible reflection [31].

Figure 7 shows the reference trace (in dashed line) and the measured trace after the event (in solid line) at wavelengths of 1310 nm and 1550 nm corresponding to a link break, a connector misalignment and a fiber bending inside the connector dead zone. In this example, the $OPL$ to the event is 15 dB, compatible with a 1:16 split-ratio and pulses of 100 ns were used. As it is expected, for a link break, a strong reflection is detected, and the estimated insertion loss at both wavelengths is very high and thus it can be considered infinite, e.g. $IL_e > 15$ dB. On the other hand, the nature of a connector misalignment can be reflective or non-reflective, and a relatively low insertion loss is normally induced. In this case, the estimated insertion loss at 1310 nm (0.21 dB) is found to be slightly larger than at 1550 nm (0.14 dB). Finally, in the case of the fiber bending, no reflection peak is detected and, as expected, the estimated event loss is much larger at 1550 nm (1.3 dB) than at 1310 nm (0.2 dB). In this case, the absence of reflection and the loss magnitude at the two wavelengths allow to classify the event as a fiber bending. Therefore, the event parameters not only can be accurately estimated with a high sensitivity using the presented method, but also the type of fault can be remotely identified even if it lies inside a dead zone. Consequently, considerable OPEX savings could be obtained.

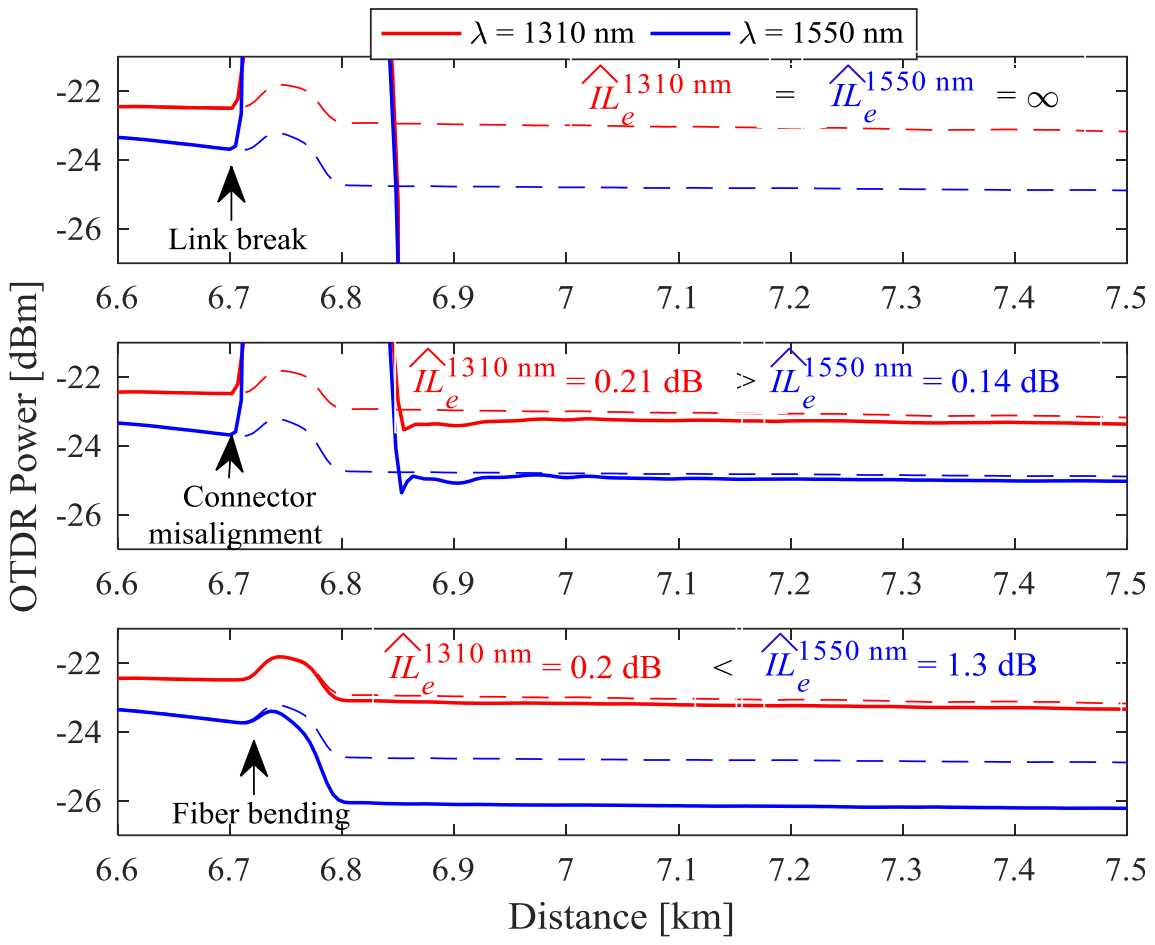

Fig. 7. Captured OTDR signals and estimated insertion loss at 1310 nm and 1550 nm for three types of fault: a link break, a connector misalignment and a fiber bending.

### C. *Estimation error*

As we have seen, the estimated insertion loss plays an essential role in the task of identifying the magnitude and type of fault. When we estimate a parameter $\theta$ by some $\hat{\theta}$, there will be a nonzero estimation error, whose magnitude is a measure of the quality of the estimate. Thus, our final analysis seek to evaluate the accuracy of the estimated event parameters. To do this, we will focus on the estimate for the insertion loss by means of Eq. (20). From this, we will not only verify how the estimation error is reduced when more observation samples are used, but also experimentally validate the theoretical model. Firstly, we define the variables

$$\begin{aligned} \widehat{IL}_e^* &= 10^{-\frac{\widehat{IL}_e}{5}} - 1 \\ IL_e^* &= 10^{-\frac{IL_e}{5}} - 1 \end{aligned} \tag{22}$$

where $\widehat{IL}_e$ is the estimate defined in Eq. (20), and therefore $\widehat{IL}_e^*$ is a random variable that follows a Gaussian distribution, since it is the sum of independent Gaussian random variables. On the other hand, $IL_e$ and therefore $IL_e^*$ are deterministic values.

We can then define the error on the estimate as

$$e = \widehat{IL}_e^* - IL_e^* \,, \tag{23}$$

which is a random variable whose PDF can be found to be

$$e \sim \mathcal{N}\left(0, \frac{\sigma_N^2}{\sum_{m=1}^{M}(OPL^2(z_m)P_0WK')^2}\right). \tag{24}$$

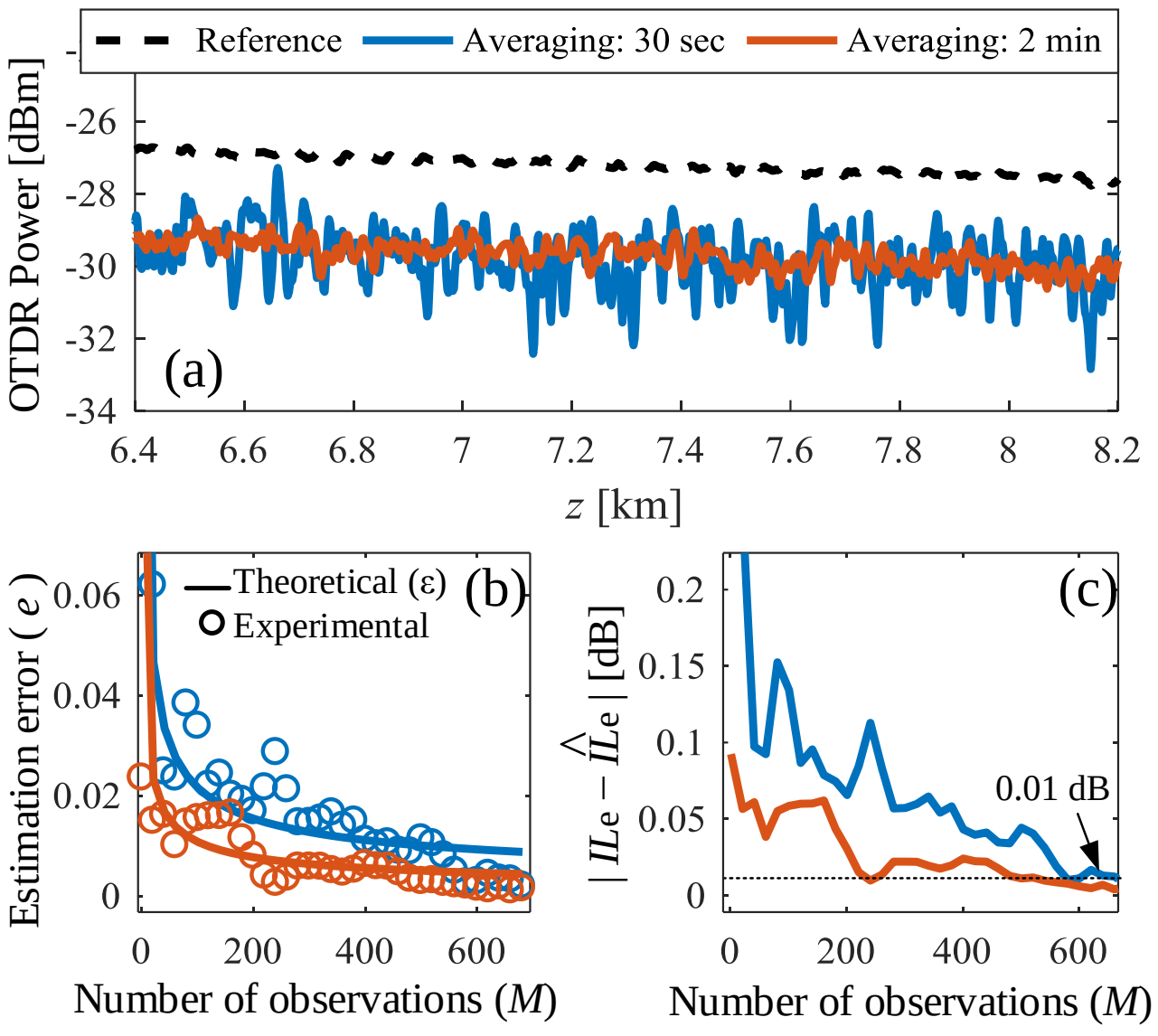

Fig. 8. Estimation error for the insertion loss: (a) samples used for the estimation, (b) theoretical and experimental error defined as in Eq. (23) and (c) experimental estimation error in dB.

To guarantee that the absolute value of this error is not higher than $\epsilon$ with probability $1-\delta$, we require that

$$\Pr\{|e| < \epsilon\} = 1 - \delta \,, \tag{25}$$

which leads to

$$\epsilon \leq \frac{\sigma_N^2}{\sum_{m=1}^{M} (OPL^2(z_m) P_0 W K')^2} Q^{-1}\left(\frac{\delta}{2}\right). \tag{26}$$

In order to compare the theoretical and experimental errors, we obtained a measurement with acquisition times of 2 minutes and 30 seconds after generating a loss inducing event with a real insertion loss of $IL_e = 1.24$ dB. In the acquired signals, which are partially shown in Fig. 8(a), the loss is detected at 680 observation samples. Thus, $M$ of the samples ($M \leq 680$) are used to estimate $IL_e$. We then analyzed the estimation error as a function of the number of observation samples $M$ used for the estimation. In Fig. 8(b), it is depicted the experimentally obtained error by means of Eq. (23) (in circles) and the theoretical value of $\epsilon$ by means of Eq. (26) (in continuous line) as a function of $M$. Here, we fixed $\delta = 0.01$. It is seen that although in some cases the experimental error is slightly higher than the theoretical upper bound $\epsilon$, it is always around this limit and following the theoretically expected trend. The slight deviations arise due to two main facts: the noise is not completely uncorrelated and the reference signal is not totally noise-free. In Fig. 8(c), it is shown the experimentally obtained absolute error $|IL_e - \widehat{IL}_e|$ in dB units, which is the error of interest. It can be seen that, for 2 minutes of averaging, using only $M = 20$ samples leads to an absolute error smaller than 0.1 dB, while for $M = 600$ samples, it is less than 0.01 dB even for short averaging times of 30 seconds. Thus, the insertion loss of low-loss inducing events can be accurately obtained with really low estimation error.

## V. Conclusion

In this work we presented DSPE-OTDR, a novel OTDR-based automatic event-detection and characterization algorithm suitable for remote monitoring of passive optical networks that provides an accurate and dead zone-free operation. The operation of DSPE-OTDR is divided into a detection stage and a parameter estimation stage.

The derivation of the detection tests are optimal according to the Neyman-Pearson criterion, thus the probabilities of detection for reflections and losses are maximized even though the event characteristic parameters are not a priori known. Our comprehensive analysis of the detection performance allows to identify the dynamic range that would be necessary to provide a desired sensitivity.

After the detection stage, the maximum-likelihood estimation of the event parameters enables to completely characterize and classify it, and therefore the type and magnitude of the fault can be remotely obtained.

In our proof-of-concept experiments we demonstrated that this method can automatically detect and characterize small events, such as a fiber bending, that lie within a dead zone, achieving high sensitivities: up to 0.14 dB in scenarios compatible with a split-ratio of 1:16 and higher than 1 dB in PONs with split-ratios up to 1:64. In addition to that, important fault parameters such as its insertion loss can be accurately estimated with minimal error, even if only a few observation samples are used for the estimation.

Since the approach operates over conventional OTDR profiles, it is completely scalable, transparent to data signals and it does not rely on the use of additional components in the ODN. In fact, only the OTDR processing software should be updated. This is extremely desirable from an operators' perspective since CAPEX and consequently OPEX are both greatly reduced compared to other monitoring solutions.

Although this solution is especially suitable for fault analysis in PON architectures, the proposed DSPE-OTDR can be equally applied to the monitoring of metropolitan and optical transport networks.

## Acknowledgment

This work was partially supported by the Universidad Nacional de Cuyo Research Projects C012, C014, C020 (3853/16), Consejo Nacional de Investigaciones Científicas y Técnicas (CONICET), Comisión Nacional de Energía Atómica (CNEA) and Sofrecom Argentina SA.